\documentstyle[12pt,a4wide]{article}
\newcommand{\be}{\begin{equation}}
\newcommand{\ee}{\end{equation}}
\newcommand{\ba}{\begin{eqnarray}}
\newcommand{\ea}{\end{eqnarray}}
\newcommand{\rlabel}[1]{\label{#1}}
\newcommand{\rref}[1]{(\ref{#1})}
\newcommand{\rcite}[1]{\cite{#1}}
\newcommand{\rbib}[1]{\bibitem{#1}}
\newcommand{\mathrm}[1]{{\rm #1}}
\newcommand{\tr}{\mathrm{tr}}

\begin{document}
\begin{titlepage}
\begin{flushright}
NORDITA - 93/69 N,P\\
NBI-93-60
\end{flushright}
\vspace{2cm}
\begin{center}
{\Large\bf ANOMALIES, VMD AND THE ENJL MODEL}\\[2cm]
{\bf Johan Bijnens$^a$ and Joaquim Prades$^{a,b,c}$}\\[0.5cm]
${}^a$ NORDITA, Blegdamsvej 17,\\
DK-2100 Copenhagen \O, Denmark\\[0.5cm]
$^b$ Niels Bohr Institute, Blegdamsvej 17,\\
DK-2100 Copenhagen \O, Denmark\end{center}
\vfill
\begin{abstract}
We show how the effective action of the
Extended Nambu--Jona-Lasinio model (ENJL) can be defined in
the presence of anomalies in a way that reproduces
the flavour anomaly. This necessarily breaks
Vector Meson Dominance (VMD) in the usual sense.
The same method can be used to construct other
chiral effective theories involving constituent quarks
and spin-1 mesons that give the correct flavour anomaly.
We also comment on how standard VMD is the root of this problem
and those observed by previous authors in implementing VMD in the
anomalous sector.
\end{abstract}
\vfill \begin{flushleft} October 1993\\
\noindent{\rule[-.3cm]{5cm}{0.02cm}} \\
\vspace*{0.2cm} \hspace*{0.5cm}
$^c$ Work supported in part by CICYT (Spain), grant Nr. AEN93-0234.
\end{flushleft} \end{titlepage}

In a recent letter Ruiz-Arriola and Salcedo\rcite{RA} claimed that the
Extended Nambu--Jona-Lasinio model (ENJL)(see refs.
\rcite{NJL,ENJL,BBR} and references therein) does not
reproduce the correct QCD
anomalous Ward identities. The correct result for the decay $\pi^0\to\gamma
\gamma$ was found but there were deviations from the anomalous Ward
identity prediction for the $\gamma\pi^0\pi^+\pi^-$ vertex. In this letter we
show how this problem can be avoided in a consistent fashion.
A similar problem was encountered in constructing anomalous effective
Lagrangians using full Vector Meson Dominance (VMD)\rcite{Brihaye}.
Our solution
also works in this case and it provides a simpler way to deal with the
Ward identities than the subtraction method used in ref. \rcite{Brihaye}.

We will first
point out the underlying cause of the problem. This followed from the way
the four quark vertices in \rcite{RA} were treated. This is essentially
equivalent to requiring VMD.
The definition of the
abnormal intrinsic parity part of the effective action for effective theories
has already quite a history. After Fujikawa derived the anomalous Ward
identities\rcite{Bardeen} from the change in the measure in the functional
integral \rcite{Fujikawa}, Bardeen and Zumino clarified the relation between
the various forms of the anomaly found using this method \rcite{BardeenZumino}.
This paper also clarified the relation between the covariant and the
noncovariant (or consistent) forms of the anomalous current.
Leutwyler then showed how these different forms are visible in the
definition of the determinant of the Dirac operator\rcite{Leutwyler}. He also
discussed the relation of the anomalous current to this determinant.
At the same time Manohar and Moore \footnote{This was done by several
independent groups but only the analysis of this reference is relevant for the
present paper.}
showed how the Wess-Zumino term\rcite{WZW}
can be derived from a change of variables
in the functional integral in a constituent chiral quark model
and how this can be used
to relate different anomalously inequivalent effective theories\rcite{Manohar}.

What we will show in this letter is that the terms that violate the
anomaly generated by the procedure used in \rcite{RA} can be subtracted
consistently. We describe how the problem with the anomaly arises in the
standard treatment of the ENJL model. Then we illustrate a simpler way
to obtain the offending terms. This way will then show that these terms
can be subtracted in a consistent fashion. We also show that our prescription
does not influence the chirally covariant part of the effective
action.

Similar problems with the anomaly occur when one tries to formulate
quark-meson effective Lagrangians which include vector and axial-vector
meson couplings to the quarks. There the problem can be solved in a similar
way by subtracting terms that contain only (axial)vector mesons and external
fields. The same basic problem also occurs when trying to implement Vector
Meson Dominance for the anomalous terms. We show how it is related to the
problem in the ENJL model and can hence be solved similarly. Finally
we explicitly state what our prescription corresponds to.

We will use the external field method to determine the physical consequences
of the underlying theory. The generating functional of the ENJL model is given
by
\ba
\rlabel{NJL}
Z(l,r,s,p)& = &\int [dq][d\bar{q}] \exp \left\{i\int d^4x\left(
\bar{q}\gamma^\mu i D_\mu q
- \bar{q}\left(s-ip\gamma_5\right)q + {\cal L}_{4-\rm{fermion}}
\right) \right\} \,;
\nonumber\\
{\cal L}_{4-\rm{fermion}} &=&
g_0 (\bar{q}^a_Lq^b_R)(\bar{q}^b_R q^a_L)
- g_1 \left[ \left( \bar{q}^a_L\gamma^\mu q^b_L\right)
\left( \bar{q}^b_L\gamma_\mu q^a_L\right) + (L\leftrightarrow R)
\right] \,;
\nonumber\\
D_\mu &\equiv&\partial_\mu -i l_\mu \gamma_L -i r_\mu\gamma_R
\ .
\ea

The inclusion of background gluonic fields can be done here as well, see
\rcite{BBR}. Here we used $\gamma_{L,R} = \left(1\pm\gamma_5\right)/2$,
$q_{L,R}=\gamma_{L,R}q$ and $\bar{q} = \left(\bar{u}\,\bar{d}\,\bar{s}\right)$.
The indices $a,b$ are flavour indices and are summed over and colour summation
within brackets is understood. The external fields $l,r,s,p$ are $3\times 3$
matrices in flavour space. The fields $l,r$ transform nonlinearly under
the chiral symmetry group $SU(3)_L\times SU(3)_R$.
The conventions used
here are those of ref. \rcite{BBR} with $g_0 = 8\pi^2 G_S/(N_c\Lambda_\chi^2)$
and $g_1 = 8\pi^2 G_V/(N_c\Lambda_\chi^2)$. $N_c$ is the number of colours.

In \rref{NJL} the measure $[dq][d\bar{q}]$ has to be defined
in a way which reproduces the correct anomalous Ward identities. This means
that the cut-off procedure should be defined with a Dirac operator that
involves the external left and right handed vector fields.

The standard way to analyze the generating functional \rref{NJL} is
to introduce
a set of auxiliary variables $M,\ L_{\mu},\ R_{\mu}$ to remove the four-fermion
couplings in the Lagrangian. This can be done by multiplying eq. \rref{NJL}
with the irrelevant constant
\ba
\rlabel{change}
&&\int[dL][dR][dM]\exp i\int d^4x \Big[
\left(
L^{ab}_{\mu}/(2\sqrt{g_1})+\sqrt{g_1}\bar{q}^b_L\gamma_\mu q^a_L \right)^2
+ (L \leftrightarrow R)
\nonumber\\&&
-\left( M^{ab}/\sqrt{g_0}+\sqrt{g_0}\bar{q}^b_L q^a_R\right)
\left( (M^\dagger)^{ba}/\sqrt{g_0}+\sqrt{g_0}\bar{q}^a_R q^b_L\right)
\Big]\ .
\ea
The fields $M$, $L_\mu$
and $R_\mu$ transform linearly under the chiral symmetry as
can be seen from \rref{change}.
We will concentrate here on the vector-axial part since it is that one that
may generate the problems with the anomaly. The part with the
scalar-pseudoscalar part is already treated in ref. \rcite{Manohar}.

The change \rref{change}
turns the four-quark terms into a ``mass'' term for the auxiliary fields
and a bilinear coupling of the auxiliary fields to the fermions.
\ba
\rlabel{NJL2}
Z(l,r,s,p)& = &\int[dM][dL][dR] \exp
i\Gamma(l,r,s,p,L,R,M) \nonumber\\&& \times
\exp i\int d^4x\left(
-\frac{1}{g_0}\tr(M^\dagger M )
+\frac{1}{4g_1}\tr\left(L_\mu L^\mu+R_\mu R^\mu\right)\right) \ .
\ea
Formally the Lagrangian in the exponential can be rewritten in terms of
the full fields $l^\prime_{\mu} = l_\mu + L_\mu$,
$r^\prime_{\mu} = r_\mu + R_\mu$ and $s^\prime,p^\prime$. The latter are
defined by $s^\prime - ip^\prime\gamma_5 = s - ip \gamma_5 + M\gamma_L
+M^\dagger \gamma_R$. We then have that
\be
\rlabel{VMD}
\Gamma(l,r,s,p,L,R,M) = \Gamma(l^\prime,r^\prime,s^\prime,p^\prime)\ .
\ee
with $\Gamma(l,r,s,p)$ defined by
\be
\exp i\Gamma(l,r,s,p) = \int [dq] [d\bar{q}] \exp\left\{i\int d^4x
\left(\bar{q} \gamma^\mu i D_\mu q -
\bar{q}\left(s-ip\gamma_5\right)q\right)\right\}\ .
\ee

We can then integrate out the fermions
to obtain the effective generating functional as a function of the external
fields and the auxiliary fields. There is one caveat here and that is
precisely the cause of the problem observed in \rcite{RA}. The measure that
corresponds to the standard procedure is then defined by a Dirac operator that
is a function of $l^\prime_\mu, r^\prime_\mu$ rather than a function of
$l_\mu$ and $r_\mu$.

Let us show in a compact fashion how this problem occurs.
For simplicity we temporarily neglect the scalar-pseudoscalar part.
The effective action $\Gamma(l^\prime,r^\prime,s,p)$ can be related
simply to $\Gamma(l,r,s,p)$ by introducing the fields
\be
l^t_\mu = l_\mu + t L_\mu\qquad\mathrm{and}\qquad r^t_\mu = r_\mu + t R_\mu\ .
\ee
These fields transform under the chiral symmetry group in the same way as
$l_\mu,r_\mu$. We can now describe the effective action after integrating
the fermions as
\begin{eqnarray}
\rlabel{llprime}
\Gamma(l^\prime,r^\prime,s,p)
 & = &
\Gamma(l,r,s,p) + \int_0^1 dt
\frac{d}{dt}\Gamma(l^t,r^t,s,p)\nonumber\\
&=& \Gamma(l,r,s,p) + \int_0^1 dt \,
\tr \left( L_\mu \frac{\delta}{\delta l^t_\mu}
\Gamma(l^t,r^t,s,p) +
 R_\mu \frac{\delta}{\delta r^t_\mu}\Gamma(l^t,r^t,s,p) \right)\ .
\end{eqnarray}
The last two terms in \rref{llprime} correspond to the left and right handed
current. This current consists out of two pieces, a
non-anomalous and an anomalous part.
The part that is non-anomalous causes no problem and one can use
the standard heat kernel methods as used in refs. \rcite{ENJL,BBR} to obtain
information about the generating functional \rref{NJL}.

The anomalous part of the current can also be written as the sum of a local
chirally invariant part and a local polynomial of ${\cal O}(p^3)$
in $l^t,r^t$\rcite{BardeenZumino}.
If we now insist that at the first step,
where we integrate out the fermions, we should have the global chiral
symmetry exact (this corresponds to choosing the left-right form of the
anomalous current)
this local polynomial contains two pieces. One is a function of
$l^t$ and its derivatives only and the other one is a function of
$r^t$ and its derivatives. This globally
invariant form is precisely the form that a ``naive''
application of the heat kernel method would give\rcite{Leutwyler}.

The anomalous left and right currents [of ${\cal O}(p^3)$] in
eq. \rref{llprime}
have the following form in the left-right symmetric scheme
\ba
\rlabel{current}
\left.\frac{\delta\Gamma(l,r,s,p)}{\delta l^{\mu}}\right|_{\rm{an}}
 & \equiv & J^L_\mu + j^L_\mu \, ; \nonumber \\
\left.\frac{\delta\Gamma(l,r,s,p)}{\delta r^{\mu}}\right|_{\rm{an}}
 & \equiv & J^R_\mu + j^R_\mu \, ; \nonumber \\
J^{L\mu} & = & \frac{N_c}{48 \pi^2}
\varepsilon^{\mu \nu \alpha \beta} \left[i
{\cal L}_\nu {\cal L}_\alpha {\cal L}_\beta +
\left\{l_{\nu \alpha} + \frac{1}{2}U^\dagger
r_{\nu \alpha} U, {\cal L}_\beta \right\} \right] \, ; \nonumber \\
j^{L\mu} & = & \frac{N_c}{48 \pi^2}
\varepsilon^{\mu \nu \alpha \beta} \left[i
l_\nu l_\alpha l_\beta +
\left\{l_{\nu \alpha}, l_\beta\right\} \right]\, ;
\nonumber\\
l(r)_{\mu\nu} & = & \partial_\mu l(r)_\nu -\partial_\nu l(r)_\mu - i
[ l(r)_\mu , l(r)_\nu]\ ,
\nonumber\\
{\cal L}_\mu & = & i U^\dagger \left(\partial_\mu U - i r_\mu U + i U l_\mu
\right)\ .
\ea
The matrix $U$ is the ``phase'' of $M$. $U = \xi^2$ with $M = \xi H \xi$.
Here $H$ is hermitian and $\xi$ is unitary. The currents $J^R_\mu$ and
$j^R_\mu$ can be obtained from $J^L_\mu$ and $j^L_\mu$ by a parity
transformation.

Since in \rref{llprime} the part $\Gamma(l,r,s,p)$ already saturates
the inhomogeneous part of the
anomalous Ward identities the remainder should be
locally chirally invariant. The
parts that are not locally invariant in the last two terms of eq.
\rref{llprime}
should thus be subtracted. As can be seen from \rref{current} these terms
are a local function of $l,L$ and their derivatives (plus the right handed
counterpart). The change in the
definition of the measure involves only the fields
$l,L,r$ and $R$ so the local terms that can be added to the effective action
to obtain the correct Ward identities should only be functions of these
and their derivatives. The preceding discussion shows that the terms that
spoil the anomalous Ward identities are precisely of this type.

As a consistency check we will show that the contribution of the
chirally invariant part of the anomalous current
to the resulting effective action can not be changed
by adding globally invariant counterterms that are functions of $l,L$ and
their derivatives only. The full list of terms that could contribute is
(an overall factor of $\varepsilon^{\mu\nu\alpha\beta}$ is understood).
\begin{eqnarray}
\tr (L_\mu L_\nu L_\alpha L_\beta) \,,&&
\tr (D_\mu L_\nu L_\alpha L_\beta)\,,
\nonumber\\
\tr (l_{\mu\nu}L_\mu L_\nu )\,,&&
\tr (l_{\mu\nu}D_\mu L_\nu)
\end{eqnarray}
and their right handed counterparts.
All others are related to these via partial integrations.
The first term vanishes because of the cyclicity of the trace. The second one
is a total derivative. The third one is forbidden by CP invariance and
the last one vanishes because of the Bianchi identities for $l_{\mu\nu}$.

This is just proving that the standard procedure of adding counterterms
and determining their finite parts by making the final effective
action satisfy the (anomalous) Ward identities also works here
giving an unambiguous answer.

We have used the left-right symmetric form of the anomaly.
But it is obvious from the discussion above, that by following an
analogous procedure to the one given here in any scheme of
regularization of the chiral anomaly one obtains the same result
since the difference between the anomalous current in two of these schemes
is a set of local polynomials that can only depend on $l_\mu'$ and
$r_\mu'$, ref. \rcite{Leutwyler}. A scheme of particular interest
is that where the vector symmetry is explicitly conserved.
In order to obtain this form of the Wess-Zumino action one
has to add a set of local polynomials that only depend on
$l'_\mu$ and $r'_\mu$
to the left-right symmetric one.
These are given explicitly in ref. \rcite{Bardeen}.

In the basis of fields we have been working until now the
effective action in the non-anomalous sector has generated a
quadratic form mixing the pseudoscalar field
and the axial-vector auxiliary fields (roughly speaking $R_\mu-L_\mu$).
It is of common practice to change to a basis where this quadratic form
is diagonal (e.g. see \rcite{BBR}).
Afterwards the vector and axial-vector degrees of freedom can be removed
by using their equations of motion to obtain an effective action for the
pseudoscalars only.
In this way one also
introduces the axial coupling, the so-called $g_A$, which in
the chiral constituent quark model \rcite{georgi}
corresponds to the axial vector coupling of constituent quarks to
pseudoscalar mesons. In our effective action, this change of basis
can only generate local chiral invariant terms that therefore
cannot modify the Wess-Zumino effective action \rcite{WZW}
and the standard predictions at order $p^4$ for $\pi^0\to
\gamma\gamma$ and $\gamma\pi^+\pi^-\pi^0$ will be satisfied. There will of
course be changes at higher orders due to the chiral invariant terms.
Thus, the value of $g_A$ is not constrained by the chiral anomaly
which is a low-energy theorem of QCD contrary to the conclusion
of ref. \rcite{RA} and in agreement with the results of ref.
\rcite{Manohar}.

These changes due to higher orders are very similar to the description using
the hidden symmetry approach\rcite{Fujiwara} (see also \rcite{Bijnens})
and the gauged Yang-Mills approach as given in
ref. \rcite{Wakamatsu}. This prescription is also
precisely the prescription that was used in ref. \rcite{ximo} to
construct the lowest order anomalous effective chiral Lagrangian
involving vector and axial-vector fields and
obtain predictions for the ``anomalous'' decays of these particles
within the ENJL model.

We would like to add one small remark about ref. \rcite{RA}. In this reference
only the $p^2$ equations of motion
for $L_\mu,R_\mu$ were used. In principle there is also
a contribution from the $p^4$ part proportional to
$\varepsilon^{\mu\nu\alpha\beta}$ when substituted into the $p^2$ term of the
effective action. This contribution does however cancel between the
$p^2$ term and the ``mass'' term for the auxiliary fields.

Our discussion was in the framework of the ENJL model. The root of the problem
was the relation \rref{VMD}. As mentioned before similar problems occur
in effective quark-meson models with explicit spin-1 mesons couplings to
the quarks and in the old approaches that require full vector meson
dominance (VMD). The basic requirement of VMD is that vector mesons couple
like the external fields. If we describe the physical
vector mesons by fields $L_\mu$
and $R_\mu$,
 this requirement can be cast in the form ($\phi$ stands for all
the other fields involved)
\ba
\rlabel{VMD2}
\left.\frac{\delta^n\Gamma(l,r,L,R,\phi)}{\delta^m L_\mu\delta^{n-m}R_\nu}
\right|_{L=R=0}
&\equiv&
\left.\frac{\delta^n\Gamma(l,r,L,R,\phi)}{\delta^m l_\mu\delta^{n-m}r_\nu}
\right|_{L=R=0}\ ; \nonumber\\*[0.5cm]
{\rm for} \,\, 0 \leq n \leq m \ .
\ea
Using the Taylor expansion
of $\Gamma(l,r,L,R,\phi)$ in $l,r,L$ and $R$ and applying
eq. \rref{VMD2} it can be shown that the action then only depends
on $l^\prime=l+L$ and $r^\prime=r+R$, i.e.
\be
 \Gamma(l,r,L,R,\phi) = \Gamma(l+L,r+R,0,0,\phi)\ .
\ee
This will lead to precisely the same type of problems as seen in the ENJL
model since this is the same relation as eq. \rref{VMD}. Here again it
can be remedied by adding local polynomials in $l,L,r$ and $R$ precisely
as was done before.

Now, what does our prescription mean in practice?
It means that vector and axial-vector fields are consistently introduced
in the low-energy effective Lagrangian by requiring a slightly
modified VMD relation
\ba
\rlabel{VMD3}
\left.\frac{\delta^n\Gamma(l,r,L,R,\phi)}{\delta^m L_\mu\delta^{n-m}R_\nu}
\right|_{L=R=0}
&\equiv&
\left(\left.
\frac{\delta^n\Gamma(l,r,L,R,\phi)}{\delta^m l_\mu\delta^{n-m}r_\nu}
\right|_{L=R=0}\right)_{\begin{array}{l}\mathrm{local}\\* \mathrm{covariant}
\end{array}}
\ ;
\nonumber\\
\mathrm{for} \,\, 1 \leq n \leq m &&
\ea
\noindent
instead of the usual VMD requirement in eq. \rref{VMD2}. This is equivalent
to use the standard heat kernel expansion technique (for a review see
\rcite{Ball}) for the non-anomalous part,i.e.
no Levi-Civita symbol, and for the chiral orders larger than $p^4$
in the anomalous part, i.e. terms with a Levi-Civita symbol.
 For the ${\cal O}(p^4)$ part of the anomalous action
one has just the usual Wess-Zumino term and for the chiral
orders smaller than $p^4$ in the anomalous part one has
\be
\int^1_0 dt \ \tr\left( L^\mu J^L_\mu(l^t,r^t,U)
+ R^\mu J^R_\mu(l^t,r^t,U) \right)\ .
\ee
Here the anomalous currents $J^{L,R}_\mu$ are
those defined in eq. \rref{current}.

In the present work we have been implicitly using a
representation similar to the
so-called vector model (model II in ref. \rcite{Ecker})
to represent vector and axial-vector fields as the
most natural way within the ENJL model we are working with.
However, it is straightforward to work out the analogous
prescription to eq. \rref{VMD3} for any other suitable
representation of vector and axial-vector fields
(tensor, gauge fields, $\cdots$)\rcite{Ecker}
to implement
VMD in both the anomalous and the non-anomalous sectors
of the effective action.

\end{document}